\documentclass[prb,preprint,amsmath,amssymb,floatfix]{revtex4}
\usepackage{graphicx}
\usepackage{natbib}
\usepackage{ulem}
\usepackage[usenames]{color}

\begin{document}

\title{Tunable Band Gaps in Bilayer Graphene-BN Heterostructures}

\author{Ashwin Ramasubramaniam}
\email{ashwin@engin.umass.edu}
\affiliation{Department of Mechanical and Industrial Engineering, University of Massachusetts Amherst, Amherst MA 01003}

\author{Doron Naveh}
\email{naveh@cmu.edu}
\author{Elias Towe}
\email{towe@cmu.edu}
\affiliation{Department of Electical and Computer Engineering, Carnegie Mellon University, Pittsburg PA 15213}

\date{\small\today}

\begin{abstract}

We investigate band-gap tuning of bilayer graphene between hexagonal boron nitride sheets, by external electric fields.  Using density functional theory, we show that the gap is continuously tunable from 0 to 0.2 eV, and is robust to stacking disorder.  Moreover, boron nitride sheets do not alter the fundamental response from that of free-standing bilayer graphene, apart from additional screening. The calculations suggest that the graphene-boron nitride heterostructures could provide  a viable route to  graphene-based electronic devices. 

\end{abstract}

%\pacs{}

\maketitle

\section{Introduction}
\label{Sec:Intro}

The nature and type of substrate on which graphene is supported  {critically influences the properties and characteristics} of  any electronic device fabricated from it.  It is generally found, for example, that commonly used SiO$_{2}$ substrates degrade the properties of pristine graphene,  resulting in significantly compromised electron  transport  and device characteristics.\cite{silica1,silica2,silica3,silica4,silica5}  While the use  of freely  suspended graphene shows superior transport properties  and impressive  device characteristics,  this form of graphene imposes several obstacles to device fabrication.  It is therefore important to explore other substrates for supporting graphene.   Recently, hexagonal boron nitride (hBN) has emerged as a potentially suitable substrate material for graphene. \cite{Dean,Usachov} Hexagonal boron nitride  is a wide gap insulator that shares similar  crystalline  structure with  graphene, but is slightly  lattice-mismatced from it by about $\sim 1.5\%$.  Micromechanically cleaved hBN layers  can generally  provide atomically smooth surfaces  with fewer charge traps and dangling bonds than the commonly used  SiO$_{2}$ surfaces.   Graphene layers on hBN have been shown to exhibit  mobilities that are about  an order of magnitude higher than those of graphene layers on SiO$_{2}$.\cite{Dean} In addition, there are theoretical studies that show the interesting prospect of  a spontaneous opening of a band gap in graphene due to the  breaking of  the A-B sublattice symmetry of  graphene on  hBN substrates. \cite{Giovanetti}  Other studies  have explored the tuning  of  band gaps in {\it single-layer} graphene on \cite{Slawinska1} and between \cite{Slawinska2} hBN sheets.  Those studies use a different approach which leads to conclusions that are different from ours.

The objective of this paper  is to report our  investigations of the possibility of tuning band gaps in bilayer graphene (BLG) supported between hBN layers -- a configuration of immediate practical relevance for electronic devices. Using density functional theory (DFT), we show that BLG essentially retains its freestanding properties\cite{Zhang,Ohta} when it is ``sandwiched''  between hBN layers; furthermore,  it  shows a tunable band gap very much like its free-standing counterpart. The tunable band gap is relatively insensitive to the stacking order of BLG relative to hBN; this is not the case for a single-layer graphene  on hBN.  Indeed, Dean {\it et al.} \cite{Dean} have noted the absence of  a band gap in single-layer graphene on hBN. They  attributed this to random stacking order, which might at best open up local gaps over short length scales. The relative insensitivity of BLG to stacking order, at least for the few cases considered here, suggests that BLG would  be a suitable candidate for graphene-on-hBN devices.  We further  consider the tunability of the  band gap of single-layer graphene  inserted between hBN layers and show, in contrast to previous  work, \cite{Slawinska2}  that single-layer graphene on hBN is a less { suitable}  candidate for electronic device applications  since it is {\it sensitive} to stacking and {\it insensitive} to applied external electric fields.

%\begin{table}[htbp]
%\caption{Band gap $E_{g}$ [eV] and gap at K-point $\Delta_{K}$ [eV] for bilayer graphene and bilayer graphene between hBN layers as a function of electric field E[V/nm].}
%\label{table}
%\begin{center}
%\begin{ruledtabular}
%\begin{tabular}{ccccc}
% E [V/nm] & \multicolumn{2}{c}{Bilayer graphene} &  \multicolumn{2}{c}{Bilayer between hBN} \\
% 	       &  $E_{g}$  & $\Delta_{K}$  & $E_{g}$  & $\Delta_{K}$ \\	       
% \cline{2-5}
% 0.0 & 0.0   & 0.0    &  0.0   &  0.0   \\
% 1.0 & 0.22 & 0.33 &  0.06 &  0.07 \\
% 2.0 & 0.28 & 0.66 &  0.15 &  0.13 \\
% 3.0 &   --    &     --   &  0.17 &  0.23 \\
% 4.0 &   --    &     --   &  0.20 &  0.30 \\
% 5.0 &   --    &     --   &  0.22 &  0.39 \\
%\end{tabular}
%\end{ruledtabular}
%\end{center}
%\label{default}
%\end{table}%

\begin{figure}[h!]
\begin{center}
\includegraphics[width=\textwidth]{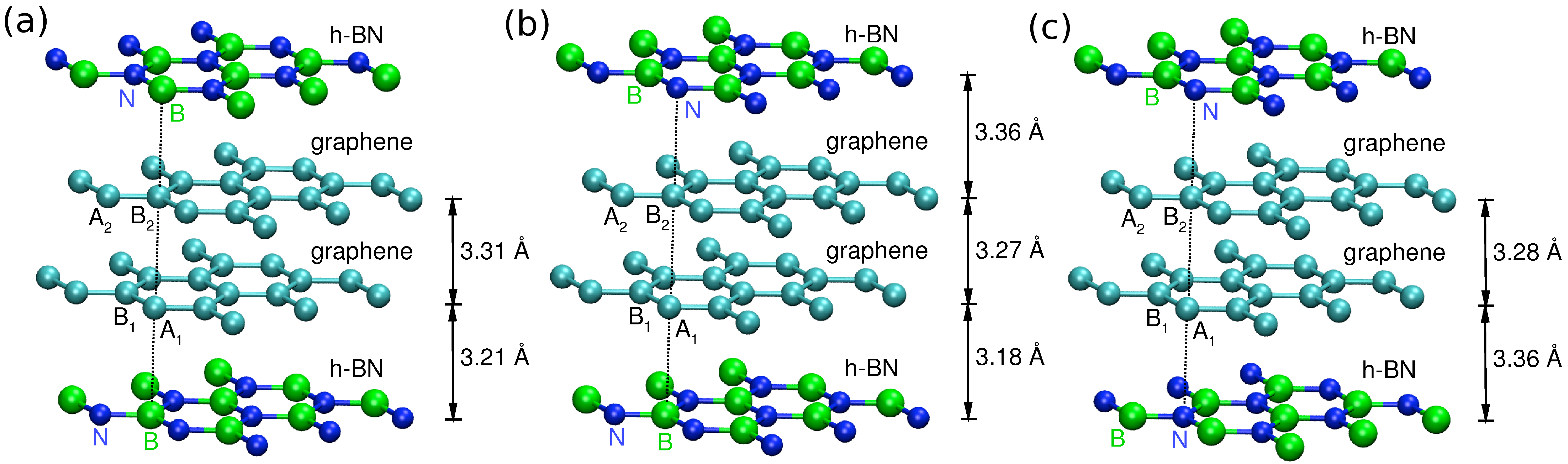}
\caption{Bilayer graphene between hBN layers. The layers have an overall Bernal (AB) stacking. Within this AB-stacking sequence the graphene and hBN layers can be arranged relative to each other to produce three different nearest-neighbor configurations along the $c$-axis, namely, (a) B-C-C-B, (b) B-C-C-N, and (c) N-C-C-N ordering. The A and B sublattices in both graphene layers are indicated in the figure. Equilibrium graphene--graphene and hBN--graphene spacings, as obtained from DFT calculations, are indicated in the figure.}
\label{Fig:quad_sch}
\end{center}
\end{figure}

\begin{figure*}[htbp]
\begin{center}
\includegraphics[width=\textwidth]{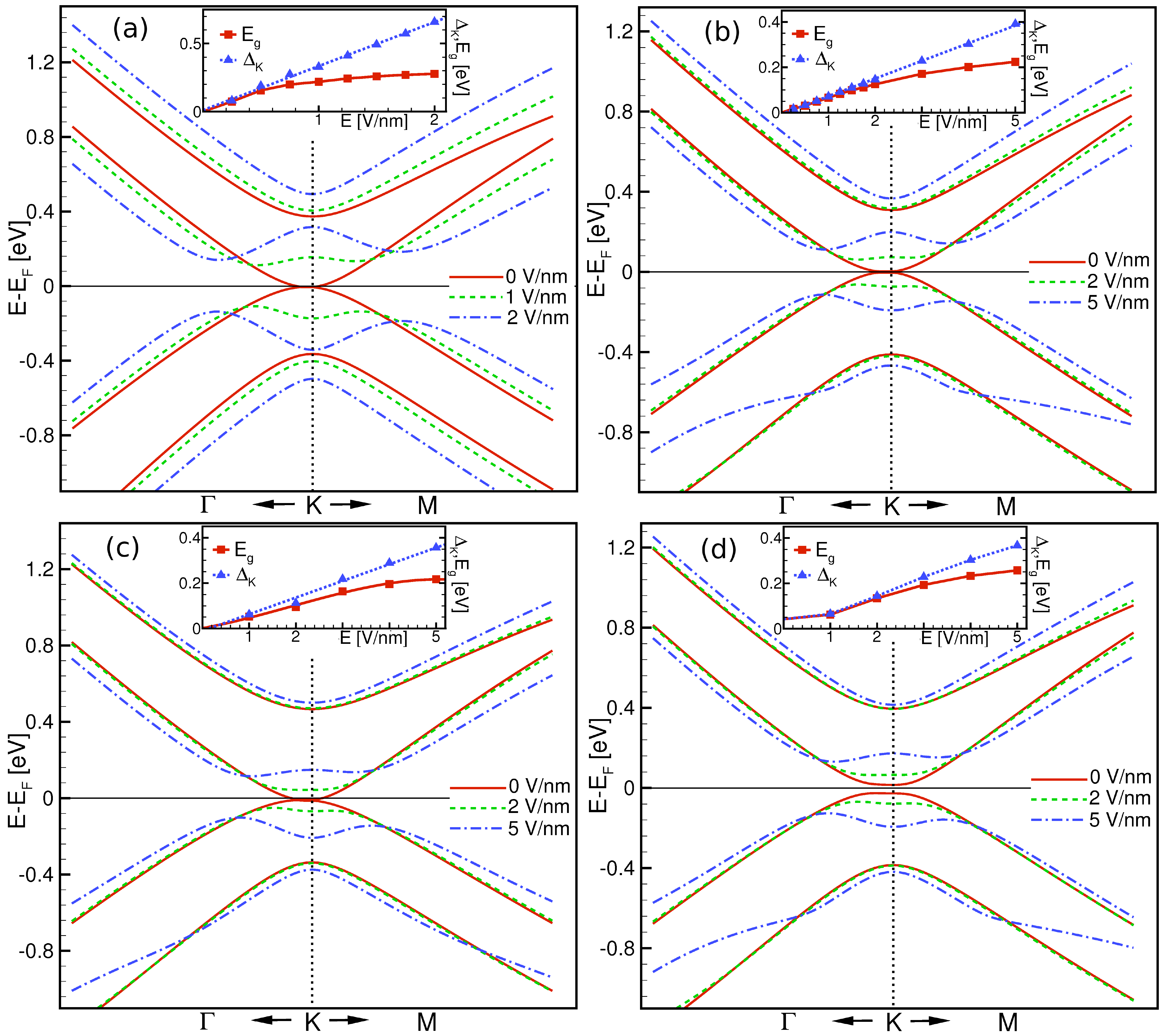}
\caption{Band structure in the vicinity of the $K$ point as a function of external electric field for (a) BLG and for BLG/hBN with (b) B-C-C-B, (c) N-C-C-N, and (d) B-C-C-N ordering of dimers along the $c$-axis. Insets show band gaps ($E_{g}$) and the gap at $K$ ($\Delta_{K}$) as a function of external field; the former saturates at higher values of E used here whereas the latter grows linearly. Here $q$ varies up to value of 0.25 around $K$, {\it i.e.,} $|K-q|\leq\pi/2a$.}
\label{Fig:gaps}
\end{center}
\end{figure*}

The structure of BLG between two hBN layers (henceforth BLG/hBN) is schematically illustrated  in Fig.\,\ref{Fig:quad_sch}. We envision that the hBN layers  play the role of  a substrate at the bottom and a top-gate dielectric for the BLG in between.  The layers are arranged in a Bernal (AB) stacking order. { In the model used for DFT simulations},  the  atomic positions and cell vectors are relaxed such that the  forces on the atoms are less than  0.01 eV/\AA~(see Methods for further details). For the stacking configurations  illustrated in Figs.\,\ref{Fig:quad_sch}(a, c) where B or N atoms are directly below the C atom of the A$_{1}$ sublattice and above the C atom of the  B$_{2}$ sublattice (forming B--C or B--N ``dimers'' in tight-binding parlance, which we adopt henceforth for convenience), there is no additional symmetry breaking arising from the relative disposition of the graphene and hBN layers. The dispersion of free-standing BLG is therefore unaltered by the presence of the substrate and one sees the usual touching of parabolic $\pi$ and $\pi^{*}$ bands at the $K$-point \cite{Geim2} [Fig.\,\ref{Fig:gaps}(b, { c})]. For the stacking sequence of  Fig.\,\ref{Fig:quad_sch}(b), in which there is a B atom below the C atom of the A$_{1}$ sublattice, and an N atom above the C atom of the  B$_{2}$ sublattice, the hBN layers break the symmetry of the BLG by inducing a dipole across the layers; this opens up a small gap ($\sim$40 meV) at the $K$ point {[Fig.\,2(d)]}. Application of an external electric  field normal to the basal planes of the structure renders the two graphene layers inequivalent, thus opening  up a band gap in the vicinity of the $K$ point while deforming the $\pi$ ($\pi^{*}$) bands such that the $K$ point is now a local minimum (maximum) surrounded by two local maxima (minima). The true band gap of the structure is then  no longer at the $K$ point but rather along the $\Gamma-K$ line of the Brillouin zone boundary.

The features described  above can be  seen in Fig.\,\ref{Fig:gaps} where we illustrate  the band structure of both freestanding BLG and BLG  sandwiched between hBN. Figure\,\ref{Fig:gaps} shows that the band structure of  the BLG/hBN system  is qualitatively identical to that of free-standing BLG.  However, there are important quantitative differences in the  actual values of band gaps.  For clarity,  the insets  in Fig.\,\ref{Fig:gaps}  show the band gaps as  functions of applied external electric  fields.  The insets show  the true band gap $(E_{g})$ and the $\pi - \pi^{*}$ band openings  at the $K$ point $(\Delta_{K})$ plotted as functions of the applied field.  The latter varies linearly with the applied external field whereas the former tends to saturate with increasing electric field.  Apparently, the hBN layers screen the BLG layers and suppress both the values of  $E_{g}$ and $\Delta_{K}$. This screening is especially evident at fields in the range of  0-2 V/nm for which the band gap for BLG/hBN system  is only about half that of  free-standing BLG.  Eventually,   both the band gap of the BLG/hBN system and that of free-standing BLG saturate at values of 0.22 eV and 0.28 eV, respectively.  In theory, this saturation is not permanent as the nature of the gap actually changes from a direct value for  the $\pi$ and $\pi^*$ bands near the  $K$-point  to an indirect one between the $\pi$ band at $K$ and the $\pi^*$ band at $\Gamma$-point;  the indirect gap progressively decreases with increasing electric field,  leading eventually to a metallic state once the $\pi^{*}$ band drops below the Fermi level at $\Gamma$. This semiconductor--metal transition is not particularly important as the fields required to reach it  are sufficiently high that they would  cause dielectric breakdown of the material  before the transition is  observed.   For more practical  fields in the range of  0--3 V/nm that  are expected to be  sustained by the BLG/hBN system, we conclude that the band gap is tunable over  a range of $\sim$0.2 eV. The key  point to note is that the dispersion of  BLG is {\it not} fundamentally altered by the presence of the hBN layers nor is there any evidence of electron or hole doping as is common on metal surfaces (Ref.\,\onlinecite{Wintterlin} and references within), SiC (Ref.\,\onlinecite{Starke} and references within) and Si/SiO$_{2}$ substrates.\cite{Romero} These  facts suggest  that hBN could be a suitable  substrate material  for BLG in device applications.

\begin{figure*}[htbp]
\begin{center}
\includegraphics[width=\textwidth]{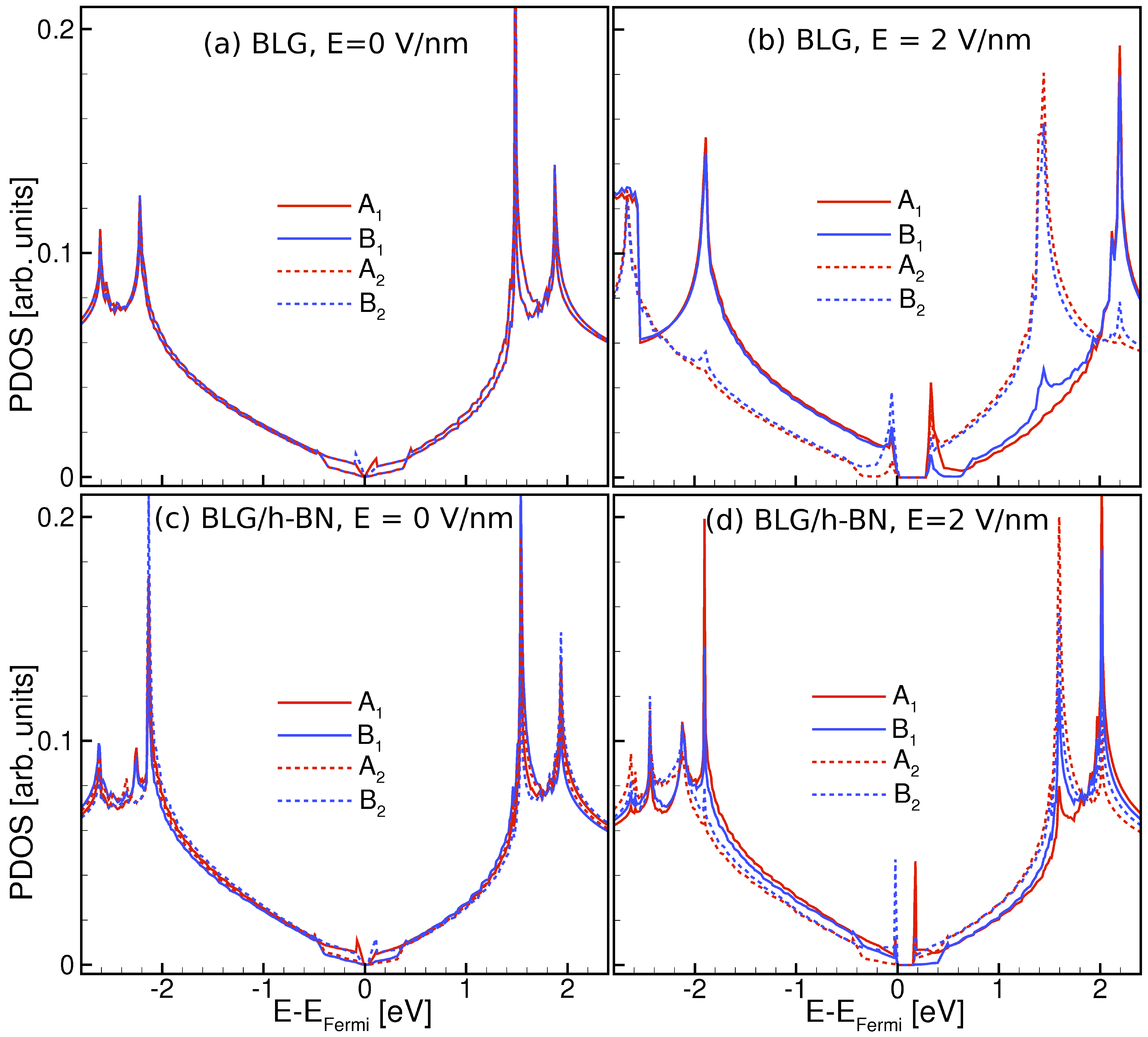}
\caption{Density of states arising from $p_{z}$ orbitals projected on individual C atoms for BLG at external fields of (a) 0 V/nm  and (b) 2 V/nm, and for BLG/hBN with B-C-C-B ordering of dimers along the $c$-axis at (c) 0 V/nm and (d) 2 V/nm. The remaining orbitals do not contribute significantly to states near the Fermi level, which are the only ones of importance here.}
\label{Fig:pdos}
\end{center}
\end{figure*}

\begin{figure*}[htbp]
\begin{center}
\includegraphics[width=\textwidth]{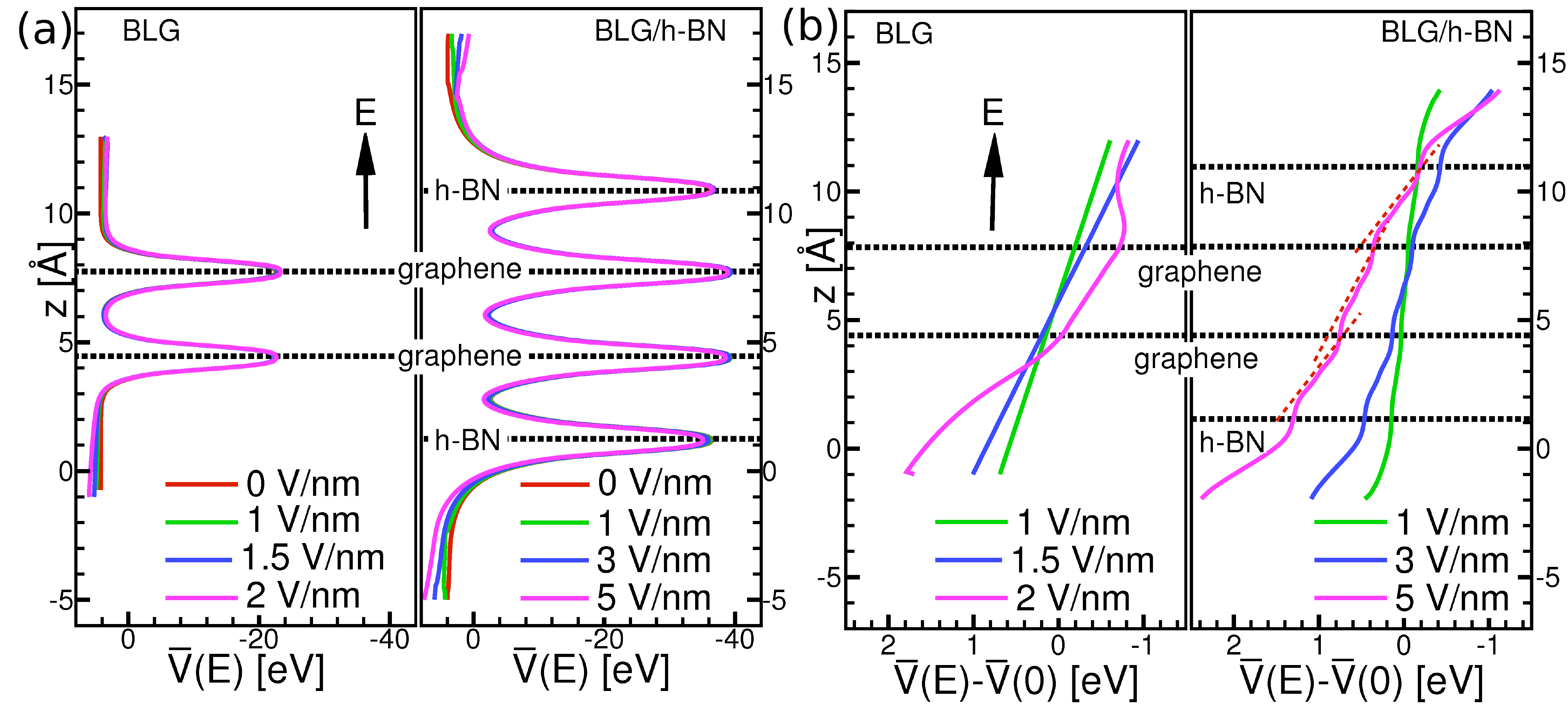}
\caption{(a) Local potential averaged over planes parallel to BLG and hBN  sheets  as a function of distance normal to the slab. (b) Difference between finite and zero-field planar averaged local potentials. Positions of the graphene and hBN planes are indicated in the figure. As seen, the hBN layers barely alter the linear variation  in local potential with external fields across the graphene planes although they provide additional screening as evidenced from the steeper slope of the planar averaged potential difference curves for BLG/hBN compared to BLG. { Also, note that the potential has a different slope between hBN and graphene and between the two graphene layers due to variations in internal screening within the structure. This is indicated by the red dotted line for the 5 V/nm case for BLG/hBN; similar variations occur for lower applied fields as well}.}
\label{Fig:locpot}
\end{center}
\end{figure*}

One can obtain further insight into the influence of external fields on the electronic structure of BLG and the  BLG/hBN system from Fig.\,\ref{Fig:pdos}, which displays the atom projected density of  states derived from the $p_{z}$ orbitals of the C atoms. These states are the only ones that contribute in the vicinity of the Fermi level and are important for understanding the electronic structure of BLG.   Figures\,\ref{Fig:pdos}(a, c) show that  there is some symmetry that is broken  between the A and B sublattices of each individual graphene layer as expected for BLG, although there is no overall symmetry breaking between the two layers themselves. This situation changes with the application of external electric fields as illustrated  in  Figs.\,\ref{Fig:pdos}(b, d), which show the asymmetric localization of valence and conduction states on each graphene layer,  thereby rendering them inequivalent. 
{ The  charge that  is  induced is mainly concentrated on the B sublattice of one layer (representing an occupied peak) whereas states at  the conduction band edge localize on the A sublattice of the other layer.\cite{Tiwari}}
For equivalent  applied fields, symmetry breaking is more noticeable for BLG than the BLG/hBN system because  the latter is screened by the hBN layers. The effect of the screening can be investigated by examining the distribution of the total local potential, which includes the usual nuclear, kinetic, Hartree, and exchange-correlation potentials,  as well as the linear potential arising from the constant external electric field.  In Fig.\,\ref{Fig:locpot}(a) we show  the planar averaged local potential as a function of distance normal to the slab structure. The same information  is displayed as the difference between the  planar averaged local potential at finite field and that at zero field in Fig.\,\ref{Fig:locpot}(b) to  clearly show  the effect of the applied field. For BLG, we see that the local potential responds linearly to applied fields of up to 1.5 V/nm; for fields larger than  2 V/nm, there is some evidence of a nonlinear response in the vacuum region near the sheet surfaces. In spite of this,  the internal screening within the graphene layers is sufficient to partially compensate the external field,  allowing  for an overall linear response of  the BLG to  external applied  fields.  This behavior  justifes the  treatment of  BLG  within the parallel plate capacitor approximation \cite{Gava,Castro2} for electric fields of this magnitude.  {  For the BLG/hBN system, we note that the response of the local potential is still linear for fields as large as 5 V/nm.  Interestingly, the difference in the planar averaged potential exhibits different slopes in the hBN--graphene and graphene--graphene regions due to different levels of internal screening within these regions [Fig.\,\ref{Fig:locpot}(b)]. Any significant nonlinearity is within the vacuum region and is an artifact attributed to proximity effects of the  dipole used to apply the external field in the (VASP) program used for the simulations.\cite{efield}  Overall, the hBN layers merely provide additional screening for the BLG; they do not fundamentally alter either the electronic properties of BLG or the tunability of its band gap.} 

\begin{figure*}[htbp]
\begin{center}
\includegraphics[width=\textwidth]{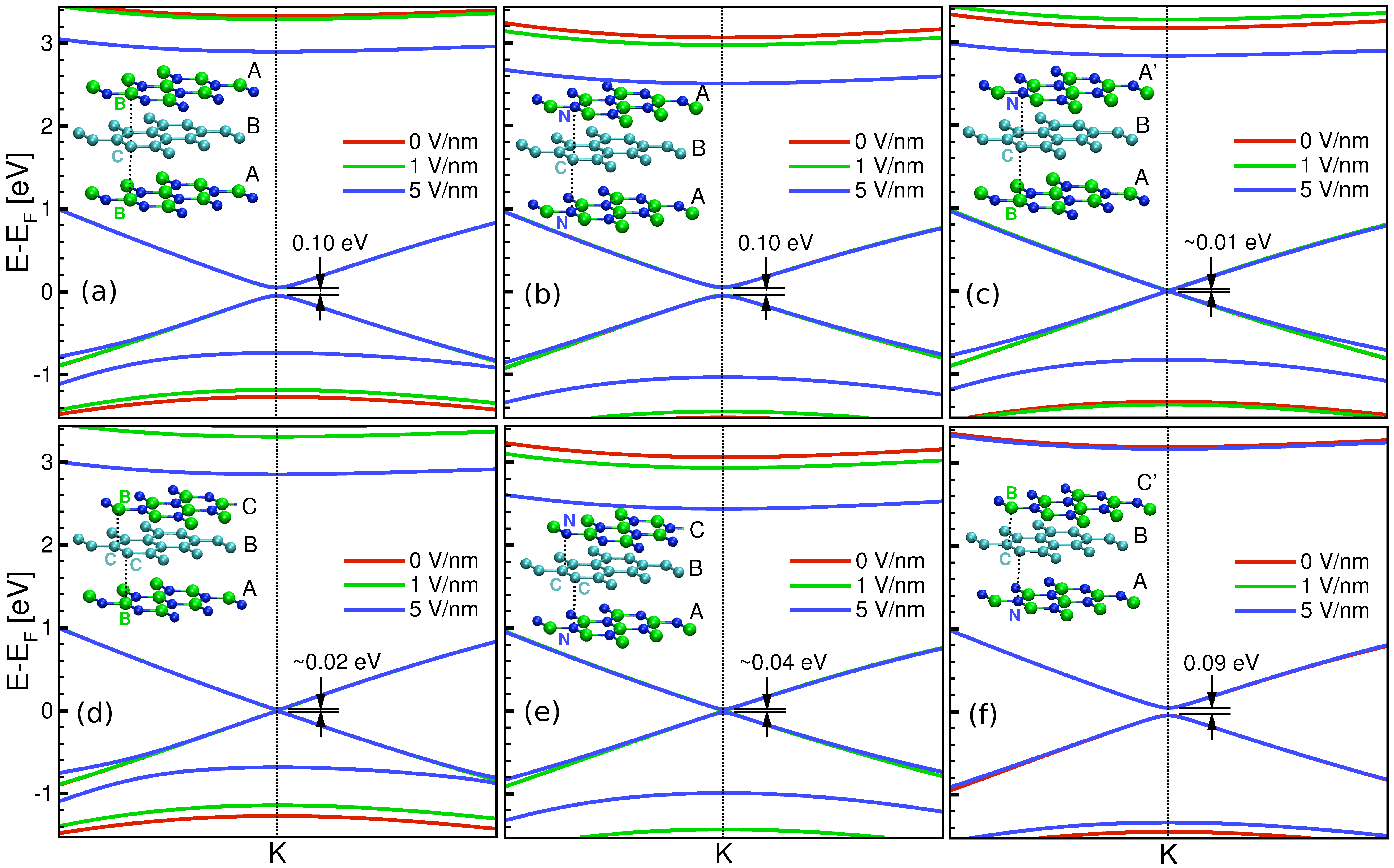}
\caption{Band structure in the vicinity of the $K$ point as a function of external electric field for single layer graphene between hBN. Insets show atomic structures for various stackings as well as the B--C and N--C dimers. The upper row consists of Bernal (AB) stacked layers while the lower row consists of rhombohedral (ABC) stackings. As seen, the band gap is sensitive to both stacking sequence as well as the dimers formed along the $c$-axis, varying from near zero gaps to a maximum of 0.1 eV. However, the gap is not particularly sensitive to external electric fields.}
\label{Fig:tri}
\end{center}
\end{figure*}

For completeness, we examine the band structure of  single-layer graphene inserted between hBN layers. It has been suggested by S\l{}awi\'nska {\it et al.}\cite{Slawinska2} that  a single layer of graphene  inserted between hBN layers in a rhombohedrally  (ABC) stacked structure  with nitrogen-carbon (N-C)  dimers can exhibit a tunable gap as large as 0.23 eV.  They further noted that other stacking confugurations  were less promising for tuning  band  gaps.  { We have repeated these calculations for  structures stacked in the Bernal and rhombohedral configurations for  all possible combinations of C--B and C--N dimers; band structures in the vicinity  of  the $K$ point are displayed Fig.\,\ref{Fig:tri} }.  Note that the usual linear dispersion characteristics of graphene are  preserved at the $K$ point,  with the presence of a band gap even at zero external fields.The magnitude of the zero-field gap is  sensitive to the stacking order and varies by about an order of magnitude (10 -- 100 meV).  However,  for applied external  fields as high as 5 V/nm,  we find no evidence of significant changes in the zero-field band-gap---the gaps are either entirely non-tunable or, at best, slightly tunable.  The difference between our results and those  of S\l{}awi\'nska {\it et al.} is likely due to the fact that they used a tight-binding model with parameters fitted to DFT calculations; their  model is probably unable to capture the physics of this system.\footnote{There is also a discrepancy between the DFT band structure in their work (Fig.\,4 of Ref.\onlinecite{Slawinska2}), which shows the unexpected presence of an energy level between the Dirac cones at the $K$ point, and the band structures reported  here.} The largest possible gap that seems to be attainable in our calculations is about $\sim 0.1$ eV {; this is not tunable, and is  substrate-induced. \cite{Giovanetti} Interestingly, the induced bandgap roughly doubles when an SLG is sandwiched between two {\it in phase} BN layers compared to when the SLG is on a BN substrate. When the two BN layers are {\it out of phase} (translated or rotated), the bandgap decays to zero}. In contrast to an SLG/hBN system, the BLG/hBN system exhibits a {\it continuously tunable} band gap of up to $\sim 0.2$ eV and is {\it relatively robust} to stacking disorder (for the few configurations considered here).

In summary, we have investigated the electronic structure of BLG inserted between hBN layers using DFT calculations.  We have examined the response of  a BLG/hBN system to external electric fields and shown that the structure  exhibits a continuously tunable band gap of up to $\sim 0.2$ eV.  We note that in general,  DFT calculations typically underestimate band gaps in materials;  so  it is conceivable that in practice larger band gaps than the ones reported in this work may be measured.  By comparing our results for the BLG/hBN system to those obtained from free-standing BLG, we have shown that the hBN layers do not fundamentally alter the electronic properties of BLG,  nor do they alter its response to electric fields, up to some screening effect.  Moreover, our results indicate that the response of the  BLG/hBN system is  fairly robust to stacking disorder, which would be an important consideration for  practical situations.  In contrast, SLG/hBN structures  are  strongly sensitive to stacking  configurations and  are virtually  insensitive to applied external electric fields.  When considered in light of reported  preliminary experiments, which  show that hBN is a  better substrate for graphene-based  devices than SiO$_{2}$,\cite{Dean} our results suggest further promising avenues for the development of electronic devices based on the BLG/hBN system.

\section*{Methods}
DFT calculations were performed using the Vienna Ab Initio Simulation Package (VASP).\cite{Vasp} Core and valence electrons were described using the Projector-Augmented Wave method.\cite{PAW1,PAW2} Electron exchange and correlation was treated using the Local Density Approximation as parameterized by Ceperley and Alder.\cite{CA} Atomic positions and cell vectors were relaxed using a conjugate gradient algorithm with a force tolerance of 0.01 eV/\AA. Electronic minimization was performed with a tolerance of $10^{-4}$ eV and electronic convergence was accelerated with a smearing of the Fermi surface by 0.05 eV. A Gaussian smearing was used during the relaxation procedure, followed by a Bl\"ochl tetrahedron smearing for accurate density of states and local potential, and a Fermi smearing for non-self-consistent band-structure calculations. A $75\times 75\times 1$ Monkhorst-Pack mesh was used for generating accurate charge densities, density of states, and local potentials. For BLG, SLG/hBN, and BLG/hBN, we used approximately 11\AA, 17\AA, and 20\AA of vacuum between periodic images of the slabs. These values were chosen to ensure a smooth vacuum-level potential. We used a 500 eV kinetic energy cutoff for SLG and BLG/hBN. A larger cutoff for 700 eV was necessary for accurate DOS and local potentials for BLG/hBN due to more vacuum  in the cell although a 500 eV cutoff was sufficient for band structures. Finally, electric fields were applied normal to the slabs in VASP, which accomplishes this by introducing dipolar sheets at the center of the simulation cell.\cite{efield}

\section*{Acknowledgments}
A. Ramasubramaniam gratefully acknowledges new faculty startup funding from the University of Massachusetts.  D. Naveh  and E. Towe  acknowledge the financial support of the Army Research Office and the National Science Foundation.

\end{document}